\documentclass[a4paper,11pt]{article}
\pdfoutput=1 

\usepackage{jheppub} 
\usepackage[T1]{fontenc} 
\usepackage[english]{babel}
\usepackage{graphicx} 
\usepackage{slashed}
\usepackage{caption}
\usepackage{amsmath}
\makeatletter
  \def\my@tag@font{\normalsize}
  \def\maketag@@@#1{\hbox{\m@th\normalfont\my@tag@font#1}}
  \let\amsmath@eqref\eqref
  \renewcommand\eqref[1]{{\let\my@tag@font\relax\amsmath@eqref{#1}}}
\makeatother
\usepackage{amsthm}
\usepackage{amsfonts}
\usepackage{mathtools}
\usepackage{subcaption}
\usepackage{tikz} 
\usepackage{listings}
\usepackage{tikz-3dplot}
\usepackage{placeins}
\usepackage{scalerel}
\usepackage{verbatim}
\usepackage{listingsutf8}
\usepackage{framed}
\usepackage{minibox}
\usepackage{float}
\usepackage{wrapfig}
\usepackage{longtable}
\usepackage{titlesec}
\usepackage[shortlabels]{enumitem}
\usepackage{multirow}
\usepackage{simpler-wick}
\usepackage{simplewick}

\usepackage{braket}
\usepackage{cases}
\usepackage{physics}
\usepackage{dsfont}
\usepackage{caption}
\usepackage{array}
\usepackage{hhline}
\usepackage{verbatim}
\usepackage{empheq}
\usepackage{mathrsfs}
\usepackage{xcolor}
\usepackage{layout}
\usepackage[mathscr]{euscript}
\usepackage{tikz-feynman}

\makeatletter
\def\@fpheader{\relax}
\makeatother

\title{Twistor Wilson loops in large-$N$ Yang-Mills theory}
\date{ }
\author[a]{Marco Bochicchio,}
\author[a,b]{Giacomo Santoni}

\affiliation[a]{Physics Department, INFN Roma1, Piazzale A. Moro 2, Roma, I-00185, Italy}
\affiliation[b]{Physics Department, Sapienza University, Piazzale A. Moro 2, Roma, I-00185, Italy}

\abstract{It has been known for many years that, in Yang-Mills theories with $\mathcal{N}=4,2,2^*$ supersymmetry, certain nontrivial supersymmetric Wilson loops exist with v.e.v. either trivial or computable by localization that arises from a cohomological field theory, which also computes the nonperturbative prepotential in $\mathcal{N}=2,2^*$ theories. Moreover, some years ago it has been argued that, in analogy with the supersymmetric case, certain nontrivial \emph{twistor Wilson loops} with trivial v.e.v. to the leading large-$N$ order exist in pure SU($N$) Yang-Mills theory and are computed, to the leading large-$N$ order, by a topological field/string theory that, to the next-to-leading $\frac{1}{N}$ order, conjecturally captures nonperturbative information on the glueball spectrum and glueball one-loop effective action as well. In fact, independently of the above, it has also been claimed that "every gauge theory with a mass gap should contain a possibly trivial topological field theory in the infrared", so that the aforementioned twistor Wilson loops realize a stronger version of this idea, as they have trivial v.e.v. at all energy scales and not only in the infrared. In the present paper, we provide a detailed proof of the triviality of the v.e.v. of twistor Wilson loops at the leading large-$N$ order in Yang-Mills theory that has previously been only sketched, opening the way to further developments.}

\begin{document}
\maketitle

\section{Introduction}

It has been known for many years that nontrivial supersymmetric (SUSY) Wilson loops with trivial vacuum expectation value (v.e.v.) exist in gauge theories with extended supersymmetry \cite{Zarembo:2002an, Guralnik:2003di, Guralnik:2004yc, Dymarsky:2006ve}. \par
Specifically, in $\mathcal{N}=4$ SUSY SU($N$) Yang-Mills (YM) theory, the following Wilson loops
\begin{equation}
\label{eq:susywilsonloop}
    W(C)=\frac{1}{N}\mathrm{tr}\ \mathrm{P}\ \mathrm{exp}\left(\oint_Ci(A_{\mu}+i\Phi_\mu)\dot{x}^\mu ds\right)
\end{equation}
have trivial v.e.v., where $A_{\mu}$ is the gauge field, $\Phi_\mu$ are four among the six scalars $\Phi_i$, and $C$ is a closed contour in the $x_1x_2$-plane. The above Wilson loops have been introduced in \cite{Zarembo:2002an}, where it has been shown that their v.e.v. is trivial at two loops in perturbation theory because of their $1/4$ supersymmetry \cite{Zarembo:2002an}. Besides, in \cite{Guralnik:2003di, Guralnik:2004yc} it has been demonstrated by means of loop equations that
\begin{equation}
    \expval{W(C)}=1
\end{equation}
to all orders of perturbation theory. Naively, triviality of the above v.e.v. arises from vast cancellations \cite{Zarembo:2002an} between the contribution of the gauge field $A_{\mu}$ and the four scalars $\Phi_\mu$, due to the factor of $i$ in front of the scalars, supersymmetry and the Euclidean invariance of the $D=10$ $\mathcal{N}=1$ SUSY YM theory, which $D=4$ $\mathcal{N}=4$ SUSY YM theory is descendant from by dimensional reduction \cite{Drukker:1999zq, Dymarsky:2006ve}. \par
SUSY Wilson loops with nontrivial v.e.v., always involving scalar fields but more general contours, have been introduced in \cite{Drukker:1999zq} and systematically computed by cohomological localization \cite{Pestun:2007rz} of the SUSY functional integral, even in theories with only $\mathcal{N}=2,2^*$ supersymmetry \cite{Pestun:2007rz}, in combination with the large-$N$ limit \cite{Russo:2019lgq}.\par
Actually, in $\mathcal{N}=2,2^*$ theories, localization of circular SUSY Wilson loops also furnishes nonperturbative information on the exact beta function and even on the exact prepotential \cite{Pestun:2007rz, Russo:2019lgq}.\par
Yet, no localization method has been presently found to compute the v.e.v. of Wilson loops in theories with less than $\mathcal{N}=2^*$ supersymmetry, let alone the corresponding nonperturbative information. \par
We may therefore wonder whether nontrivial Wilson loops with trivial v.e.v. exist in pure Yang-Mills theory.\par
Some years ago, it has been pointed out that the large-$N$ limit of pure SU($N$) Yang-Mills theory admits a class of \emph{twistor Wilson loops} \footnote{The name originates from the parameter $\lambda$ -- entering the gauge connection by which twistor Wilson loops are defined in eq. \eqref{twl00} -- that has been interpreted \cite{Bochicchio:2012bj} as the fiber of a twistor fibration over spacetime.} -- introduced in \cite{Bochicchio:2012bj} -- with trivial v.e.v. to the leading large-$N$ order and to all orders in the 't Hooft coupling $g$ in perturbation theory \cite{Bochicchio:2012bj}. Naively, here triviality arises from vast cancellations (section \ref{sec:twistwl}) that occur by a mechanism similar to the SUSY case, involving $D=4$ Euclidean invariance and the large-$N$ limit (instead of supersymmetry).\par
In fact, independently of \cite{Bochicchio:2012bj}, it has been claimed that "every gauge theory with a mass gap should contain a possibly trivial topological field theory (TFT) in the IR"\footnote{See footnote g, p. 9 in \cite{Bochicchio:2012bj} and \cite{SCGP}.}. \par
Actually, the twistor Wilson loops in \cite{Bochicchio:2012bj} constitute a trivial TFT at all energy scales underlying the large-$N$ limit of pure YM theory -- not only in the IR -- since they define trivial homology invariants of planar \footnote{Here planarity refers to the topology of the Feynman diagrams contributing to the leading large-$N$ order to gauge-invariant observables both in U($N$) and SU($N$) YM theory, according to 't Hooft \cite{tHooft:1973alw}.} YM theory because of the shape independence of their v.e.v. to the leading large-$N$ order. \par
The last observation is fundamental for future developments because it has been argued in \cite{Bochicchio:2016toi} that the above TFT in large-$N$ YM theory \cite{Bochicchio:2012bj} can be realized, to the leading large-$N$ order, by a version of Chern-Simons theory \cite{Bochicchio:2016toi} on noncommutative spacetime \cite{Minwalla:1999px,Szabo:2001kg}.
\par The aforementioned TFT can be extended to the next-to-leading large-$N$ order by coupling Chern-Simons to string D-branes \cite{Bochicchio:2016toi}, in a way that conjecturally captures nonperturbative information \cite{Bochicchio:2016toi} on the glueball spectrum and the glueball one-loop effective action -- somehow in analogy with the prepotential in the aforementioned cohomological $\mathcal{N}=2^*$ theory -- but rather by employing homological methods \cite{Bochicchio:2012bj, Bochicchio:2016toi}. \par
This program has received a recent revival by the suggestion \footnote{See footnote 7 p. 9 in \cite{Bochicchio:2024obj}.} that the nonperturbative would-be glueball one-loop effective action in \cite{Bochicchio:2016toi} should be interpreted as the generating functional of certain correlators of twist-$2$ operators with the topology of pinched tori \cite{Bochicchio:2024obj} -- instead of as the corresponding generating functional of the nonperturbative one-loop collinear $S$ matrix -- that opens the way to further developments \cite{Bochicchio2025}.\par
Because of its intrinsic interest and in preparation for the aforementioned nonperturbative developments, we provide in the present paper a detailed proof of the triviality to the leading large-$N$ order and to all orders in $g$ of the v.e.v. of twistor Wilson loops in pure YM theory -- which has been only sketched in \cite{Bochicchio:2012bj} -- along lines that we describe as follows. \par
Twistor Wilson loops \cite{Bochicchio:2012bj} are constructed by means of a noncommutative deformation of ordinary (Euclidean) YM theory \cite{Szabo:2001kg, Minwalla:1999px}, where the four spacetime (complex) coordinates $w,\bar{w}, u, \bar{u}$ are promoted to operators $\hat w,\hat{\bar{w}}, \hat u, \hat{\bar{u}}$ satisfying the commutation relations
\begin{equation}
\label{eq:vartheta}
    \left[\hat{u},\hat{\bar{u}}\right]=\vartheta_1\hat{1}\ , \qquad \left[\hat{w},\hat{\bar{w}}\right]=\vartheta_2\hat{1}  \ ,
\end{equation}
with $\vartheta_1, \vartheta_2$ constants. Hence, their construction requires a short detour into gauge theories on noncommutative spacetime -- for short, noncommutative gauge theories. \par
The large-$N$ 't Hooft expansion \cite{tHooft:1973alw} also applies \cite{Minwalla:1999px,Szabo:2001kg} to U($N$) noncommutative pure Yang-Mills theory (and, incidentally, to its SUSY extensions), so that the corresponding large-$N$ notion of planarity holds (appendix \ref{app:ncfields}). \par
As long as we are interested in correlators of elementary fields $\phi_i(p_i)$ in the momentum representation, the planar limit of the theory on commutative spacetime can be reconstructed from the planar limit (appendix \ref{app:ncfields}) of its noncommutative counterpart, defined on noncommutative spacetime (section \ref{3})
\begin{equation}
\label{eq:ncalg1}
    \left[\hat{x}^{\mu},\hat{x}^{\nu}\right]=i\theta^{\mu\nu}\hat{1} \ ,
\end{equation}
by simply taking the limit of zero noncommutativity $\theta^{\mu\nu} \rightarrow 0$, according with the following fundamental result \cite{Filk:1996dm}. \par
Denoting as $\expval{\dots }^{(\theta)}_{\text{conn, pl}}$ the connected planar v.e.v. in the \emph{noncommutative} theory and as $\expval{\dots }_{\text{conn, pl}}$ the connected planar v.e.v. in the \emph{commutative} theory, we get \cite{Filk:1996dm}
\begin{equation}
    \expval{ {\phi}_{i_1}(p_1)\dots {\phi}_{i_n}(p_n)}^{(\theta)}_{\text{conn, pl}}=e^{-\frac{i}{2}\sum_{i<j}p_i\wedge p_j}\expval{{\phi}_{i_1}(p_1)\dots {\phi}_{i_n}(p_n)}_{\text{conn, pl}} \ ,
\end{equation}
where, for any two vectors $v$ and $w$, we define
\begin{equation}
	v\wedge w = \theta^{\mu\nu}v_{\mu}w_{\nu} \, .
\end{equation}
This reconstruction procedure is not applicable to nonplanar diagrams because, in this case, the limit of zero noncommutativity is singular \cite{Minwalla:1999px,Szabo:2001kg}. \par
Besides, in noncommutative gauge theories the gauge-invariant observables are nonlocal \cite{Gross:2000ba,Szabo:2001kg}. They are obtained by dressing with a Wilson line local operators $O(x)$ transforming in the adjoint representation of the gauge group and integrating over all spacetime \cite{Gross:2000ba,Szabo:2001kg} (section \ref{sec:obs})
\begin{equation}
\label{eq:gaugeinv0}
    \widetilde{O}=\frac{1}{N}\mathrm{tr}\int d^Dx\  O(x)* \mathrm{P}\ \mathrm{exp}_*\left(i\int_{C_v}A_{\mu}(x+\xi)d\xi^{\mu}\right)* e^{iv_{\mu}(\theta^{-1})^{\mu\nu}x_{\nu}}\ , 
\end{equation}
where $v=\xi(1)-\xi(0)$, $*$ denotes the Groenewold-Moyal product that is employed to define a realization (section \ref{3}) of the operator algebra in eq. \eqref{eq:ncalg1}. \par
For such operators, in the quantum theory, the limit of zero noncommutativity is generally ill defined even in the planar sector, due to new singularities appearing because of the contour integration \cite{Gross:2000ba}. The same considerations apply in general to noncommutative Wilson loops, which are defined by simply setting $O(x)=1$ and $v=0$ in eq. \eqref{eq:gaugeinv0}.
\par 
After the above preliminaries, twistor Wilson loops are defined as (section \ref{sec:twistwl})
\begin{align} \label{twl00}
    &W_{\lambda}(C_{u\bar{u}})= \nonumber \\ 
    &\frac{1}{N\mathrm{Tr}'(\hat{1})}  \mathrm{Tr}'\int \frac{d^2u}{V_2}\ \mathrm{tr}\ \mathrm{P}\ \mathrm{exp}_{*'}\Big[i\oint_{C_{u\bar{u}}}(\hat{A}_u(u+\zeta,\bar{u}+\bar{\zeta},\hat{w},\hat{\bar{w}})+\lambda \hat{D}_w(u+\zeta,\bar{u}+\bar{\zeta},\hat{w},\hat{\bar{w}}))d\zeta \nonumber \\
    +&(\hat{A}_{\bar{u}}(u+\zeta,\bar{u}+\bar{\zeta},\hat{w},\hat{\bar{w}})+\lambda^{-1} \hat{D}_{\bar{w}}(u+\zeta,\bar{u}+\bar{\zeta},\hat{w},\hat{\bar{w}}))d\bar{\zeta}\Big] \ ,
\end{align}
where $\vartheta=\vartheta_1=\vartheta_2$, $\lambda$ is a nonzero complex parameter, $C_{u\bar{u}}$ is a closed contour lying on the plane $u\bar{u}$, $*'$ denotes the Groenewold-Moyal product restricted to the coordinates $u,\bar{u}$, $\mathrm{Tr}'$ denotes the trace over the Fock space on which the noncommutative coordinates $\hat{w}, \hat{\bar{w}}$ are represented (appendix \ref{app:ncspaces}), $V_2$ is the volume of two-dimensional spacetime, and the normalization factors are chosen so that that eq. \eqref{eq:trivial0} holds. \par
The aim of the present paper is to provide a detailed proof of the statement
\begin{equation}
	\label{eq:trivial0}
    \lim\limits_{N\to +\infty} \expval{W_{\lambda}(C_{u\bar{u}})} = 1
\end{equation}
for any value of $\vartheta$ and to all orders in $g$. In fact, the triviality of the v.e.v. in the above equation arises before integrating on the contour of the loop, provided that we choose a certain intermediate regularization (appendix \ref{C}). \par
Hence, twistor Wilson loops are well defined in the planar limit, i.e. to the leading large-$N$ order and, as a consequence, in the commutative limit $\vartheta \rightarrow 0$ \emph{after} taking the large-$N$ limit. Thanks to their trivial v.e.v. regardless of the shape of the contour, they define trivial homology invariants of planar YM theory.

\section{Plan of the paper}

In section \ref{3} we define the algebra of noncommutative spacetime. \par
In section \ref{sec:ncym} we define noncommutative YM theory in both the coordinate and operator representation. \par
In section \ref{sec:obs} we define the gauge-invariant observables of noncommutative gauge theories, proving the equivalence of two different definitions in the literature. \par
In section \ref{6} we compute noncommutative Wilson loops to the leading large-$N$ order in terms of the corresponding commutative planar objects. \par
In section \ref{sec:twistwl} we recall the definition of twistor Wilson loops both in the operator and coordinate representations and demonstrate the triviality of their v.e.v. in the planar limit. \par
In section \ref{conclusion} we state our conclusions.\par
In appendix \ref{app:ncspaces} we define the algebra of noncommutative spacetime and its representations. \par
In appendix \ref{app:ncfields} we review the basics of quantum field theories on noncommutative spacetime and, specifically, the notion of planarity in noncommutative theories.\par
In appendic \ref{C} we introduce a suitable intermediate regularization.

\section{Algebra of noncommutative spacetime} \label{3}

Following \cite{Szabo:2001kg}, we define the $D$-dimensional (Euclidean) noncommutative spacetime $\mathbb{R}^D_{\theta}$ by the algebra
\begin{equation}
\label{eq:ncalg}
    \left[\hat{x}^{\mu},\hat{x}^{\nu}\right]=i\theta^{\mu\nu}\hat{1} \ ,
\end{equation}
with the spacetime dimension $D$ even and $\theta^{\mu\nu}$ invertible. We also define 
\begin{equation}
	\label{eq:partial}
	\hat{\partial}_{\mu}(\hat{O}) =\left[-i(\theta^{-1}\hat{x})_{\mu},\hat{O} \right] \ .
\end{equation}
$\hat{\partial}_{\mu}(\cdot)$ can be extended as an inner derivation of the algebra in eq. \eqref{eq:ncalg} satisfying the commutation relations
\begin{align}
	\label{eq:partialcomm}
    & \left[\hat{\partial}_{\mu},\hat{\partial}_{\nu}\right]=0 \ ,&& \left[\hat{\partial}_{\mu}, \hat{x}^{\nu}\right]={\delta_{\mu}}^{\nu}\hat{1} \ .
\end{align}
The algebra generated by $\hat{x}_{\mu}$, together with its inner derivations $\hat{\partial}_{\mu}$, admits a Fock representation $\mathcal{H}$ (appendix \ref{app:ncspaces}). The trace Tr over $\mathcal{H}$ satisfies (appendix \ref{app:trace})
\begin{equation}
\label{eq:pf}
    (2\pi)^{D/2}\mathrm{Pf}(\theta)\mathrm{Tr}[e^{ik_{\mu}\hat{x}^{\mu}}]=(2\pi)^{D}\delta^{(D)}(k) \ .
\end{equation}
To a function $f: \mathbb{R}^D\longrightarrow \mathbb{R}^D$ that decays sufficiently fast at infinity, we associate the operator-valued function $\hat{f}(\hat{x})$
\begin{equation}
    \hat{f}(\hat{x})\equiv\int\frac{d^Dk}{(2\pi)^D}\int d^Dx\ f(x)e^{ik_{\nu} x^{\nu}}e^{-ik_{\rho}\hat{x}^{\rho}} \ .
\end{equation}
From eqs. \eqref{eq:pf} and \eqref{eq:expidentity} it follows
\begin{equation}
    (2\pi)^{D/2}\mathrm{Pf}(\theta)\mathrm{Tr}\left[\hat{f}_1(\hat{x})\ \dots \ \hat{f}_n(\hat{x})\right]=\int d^Dx\ f_1(x)*\dots * f_n(x) \ ,
\end{equation}
where $*$ denotes the Groenewold-Moyal product
\begin{equation}
    f_1(x)*\dots *f_n(x)=\prod_{j<k}^n\mathrm{exp}\left(\frac{i}{2}\theta^{\mu\nu}\frac{\partial}{\partial x_j^{\mu}}\frac{\partial}{\partial x_k^{\nu}}\right)f_1(x_1)\dots f_n(x_n)\Bigg\rvert_{x_1=\dots =x_n=x} \ ,
\end{equation}
whose definition at noncoinciding points is entirely analogous to the one above. The Groenewold-Moyal product provides a coordinate representation of the spacetime algebra in eq. \eqref{eq:ncalg}
\begin{equation}
x^{\mu} * x^{\nu} -x^{\nu} * x^{\mu} = i\theta^{\mu\nu}1 \ .
\end{equation}
Remarkably,
\begin{align}
	\label{eq:twostar}
	\int d^Dx\ f_1(x)*f_2(x)=\int d^Dx\ f_1(x)f_2(x) \ .
\end{align}
The above identity follows by integrating by parts and using the antisymmetry of the matrix $\theta_{\mu\nu}$. Incidentally, this property implies that in a noncommutative field theory the propagators are equal to their commutative counterparts (appendix \ref{app:plan}).
\par 
A consequence of the commutation relations in eq. \eqref{eq:ncalg} is that
\begin{align}
	\label{eq:thetax}
	&\left[-i(\theta^{-1}\hat{x})_{\mu}, \hat{f}(\hat{x})\right]=\widehat{(\partial_{\mu}f)}(\hat{x}) \ .
\end{align}
We first prove it for the exponential function $e^{-ik_{\rho}\hat{x}^{\rho}}$
\begin{align}
	\left[-i(\theta^{-1}\hat{x})_{\mu}, e^{-ik_{\rho}\hat{x}^{\rho}}\right]=&\sum_{n=0}^{\infty}\frac{(-i)^n}{n!}k_{\mu_1}\dots k_{\mu_n}\left[-i(\theta^{-1}\hat{x})_{\mu},\hat{x}^{\mu_1}\dots \hat{x}^{\mu_n}\right] \nonumber \\
	=&\sum_{n=0}^{\infty}\frac{(-i)^n}{n!}k_{\mu_1}\dots k_{\mu_n}\sum_{k=1}^{n}\hat{x}^{\mu_1}\dots\left[-i(\theta^{-1}\hat{x})_{\mu}, \hat{x}^{\mu_k}\right]\dots\hat{x}^{\mu_n} \nonumber \\
	=&-ik_{\mu}e^{-ik\hat{x}} \ ,
\end{align}
where in the second line the commutator acts as an inner derivation and in the third line we have employed eq. \eqref{eq:ncalg}. Then, it follows for a generic function $\hat{f}(\hat{x})$
\begin{align}
	\label{eq:expcomm}
	\left[-i(\theta^{-1}\hat{x})_{\mu}, \hat{f}(\hat{x})\right]=&\int\frac{d^Dk}{(2\pi)^D}\int d^Dx\ f(x)e^{ik_{\nu} x^{\nu}}\left[-i(\theta^{-1}\hat{x})_{\mu},e^{-ik_{\rho}\hat{x}^{\rho}}\right] \nonumber \\
	=&\int\frac{d^Dk}{(2\pi)^D}\int d^Dx\ f(x)e^{ik_{\nu} x^{\nu}}(-ik_{\mu})e^{-ik_{\rho}\hat{x}^{\rho}} \nonumber \\
	=&\int\frac{d^Dk}{(2\pi)^D}\int d^Dx\ f(x)\left(-\partial_{\mu}e^{ik_{\nu} x^{\nu}}\right)e^{-ik_{\rho}\hat{x}^{\rho}} \nonumber \\
	=&\int\frac{d^Dk}{(2\pi)^D}\int d^Dx\ \partial_{\mu}f(x)e^{ik_{\nu} x^{\nu}}e^{-ik_{\rho}\hat{x}^{\rho}} \nonumber \\
	=&\widehat{(\partial_{\mu}f)}(\hat{x}) \ ,
\end{align}
where in the fourth line we have integrated by parts.\par
Remarkably, in noncommutative gauge theories translations can be realized by gauge rotations
\begin{align}
\label{eq:transl}
    e^{-ia_{\mu}(\theta^{-1}\hat{x})^{\mu}}\hat{f}(\hat{x})e^{+ia_{\mu}(\theta^{-1}\hat{x})^{\mu}}=&\sum_{n=0}^{\infty}\frac{1}{n!}\left[\left\{-ia\theta^{-1}\hat{x}\right\}^n,\left\{\hat{f}(\hat{x})\right\}\right] \nonumber \\
    =&\sum_{n=0}^{\infty}\frac{1}{n!}a_{\mu_1}\dots a_{\mu_n}\widehat{(\partial^{\mu_1}\dots\partial^{\mu_n}f)}(\hat{x}) \nonumber \\
    =&\hat{f}(\hat{x}+a) \ ,
\end{align}
where we have employed \cite{Volkin1968}
\begin{align}
	e^ABe^{-A}=\sum_{n=0}^{\infty}\frac{1}{n!}\left[\{A\}^n,\{B\}\right] \ ,
\end{align}
with the symbol
\begin{align}
	\left[\{A\}^n,\{B\}\right]=[\underbrace{A,\dots,[A,[A}\limits_{n\ \text{times}},B]]] \ .
\end{align}
In the coordinate representation eq. \eqref{eq:transl} reads
\begin{equation}
\label{eq:translcoord}
    e^{-ia_{\mu}(\theta^{-1})^{\mu\nu}x_{\mu}}*f(x)*e^{+ia_{\mu}(\theta^{-1})^{\mu\nu}x_{\nu}}=f(x+a) \ .
\end{equation}
Similarly, if $\hat{\partial}_{\mu}$ is interpreted as a derivation, we get
\begin{align}
	\label{eq:partialfhat}
	\left[\hat{\partial}_{\mu}, \hat{f}(\hat{x})\right]=\widehat{\partial_{\mu}f}(\hat{x})
\end{align}
The proof is identical to the proof of eq. \eqref{eq:thetax}, except for the fact that we employ eq. \eqref{eq:partialcomm} instead of eq. \eqref{eq:ncalg}. As a consequence, we have
\begin{align}
	\label{eq:transl1}
	e^{a^{\mu}\hat{\partial}_{\mu}}\hat{f}(\hat{x})e^{-a^{\mu}\hat{\partial}_{\mu}}=&\sum_{n=0}^{\infty}\frac{1}{n!}[\{a\hat{\partial}\}^n,\left\{\hat{f}(\hat{x})\right\}] \nonumber \\
	=& \sum_{n=0}^{\infty}\frac{1}{n!}a_{\mu_1}\dots a_{\mu_n}\widehat{(\partial^{\mu_1}\dots\partial^{\mu_n}f)}(\hat{x})\nonumber \\
	=&\hat{f}(\hat{x}+a) \ ,
\end{align}
where we have repeatedly employed the second commutation relation in eq. \eqref{eq:partialcomm}.

\section{Noncommutative Yang-Mills theory}
\label{sec:ncym}

The elementary field of U($N$) Yang-Mills theory on $\mathbb{R}^D_{\theta}$  \cite{Szabo:2001kg} is the $\mathfrak{u}(N)$-valued connection $A_{\mu}=A_{\mu}^at^a$, where $t^a$ are the generators of the Lie algebra $\mathfrak{u}(N)$ in the fundamental representation. The gauge symmetry is defined as
\begin{equation} \label{gg}
    A_{\mu}(x)\longmapsto g(x)* A_{\mu}(x)* g^{-1}(x)-ig(x)*\partial_{\mu}g^{-1}(x) \ ,
\end{equation}
where $g(x)\in$ U($N$) and $g^{\dagger}(x)*g(x)=g(x)*g^{\dagger}(x)=\mathds{1}_N$. The corresponding field strength tensor is
\begin{equation}
    F_{\mu\nu}=\partial_{\mu}A_{\nu}-\partial_{\nu}A_{\mu}-i\left(A_{\mu}* A_{\nu}-A_{\nu}* A_{\mu}\right) \ ,
\end{equation}
which transforms as $F_{\mu\nu}\longmapsto g* F_{\mu\nu}* g^{-1}$, i.e. in the adjoint representation of the gauge group. The Yang-Mills action is thus
\begin{equation}
\label{eq:ymaction}
    S_{YM}=\frac{N}{2g^2}\int d^Dx\ \mathrm{tr}\left(F_{\mu\nu}* F_{\mu\nu}\right) \ ,
\end{equation}
which is clearly gauge invariant, with $g$ the 't Hooft coupling and $\mathrm{tr}$ the matrix trace for $\mathfrak{u}(N)$. The above action can also be written in the operator representation. To this aim, we define the two quantities \cite{Douglas:2001ba, Szabo:2001kg}
\begin{equation}
\label{eq:cd}
    \hat{C}_{\mu}=-i(\theta^{-1}\hat{x})_{\mu}+i\hat{A}_{\mu}(\hat{x}) \ , \qquad \hat{D}_{\mu}=\hat{\partial}_{\mu}+i\hat{A}_{\mu}(\hat{x})  \ .
\end{equation}
It follows from eq. \eqref{gg} that in the operator representation the connection $\hat{A}_{\mu}(\hat{x}) $ transforms as
\begin{align}
	\hat{A}_{\mu}(\hat{x}) \longmapsto \hat{g}(\hat{x}) \hat{A}_{\mu}(\hat{x}) \hat{g}^{-1}(\hat{x}) -i\hat{g}(\hat{x}) \left(\widehat{\partial_{\mu}g^{-1}}\right)(\hat{x}) \ .
\end{align}
From eq. \eqref{eq:thetax} it follows that $\hat{C}_{\mu}$ transforms in the adjoint representation of the gauge group,
\begin{align}
	\hat{C}_{\mu}\longmapsto & -i(\theta^{-1}\hat{x})_{\mu}+i\hat{g}(\hat{x}) \hat{A}_{\mu}(\hat{x}) \hat{g}^{-1}(\hat{x}) +\hat{g}(\hat{x}) \left(\widehat{\partial_{\mu}g^{-1}}\right)(\hat{x}) \nonumber \\
	=& -i(\theta^{-1}\hat{x})_{\mu}+i\hat{g}(\hat{x}) \hat{A}_{\mu}(\hat{x}) \hat{g}^{-1}(\hat{x}) +\hat{g}(\hat{x}) \left[-i(\theta^{-1}\hat{x})_{\mu},\hat{g}^{-1}(\hat{x})\right] \nonumber \\
	=&\hat{g}(\hat{x})\left(-i(\theta^{-1}\hat{x})_{\mu}+i\hat{A}_{\mu}(\hat{x})\right)\hat{g}^{-1}(\hat{x}) \nonumber \\
	=&\hat{g}(\hat{x})\hat{C}_{\mu}\hat{g}^{-1}(\hat{x}) \ .
\end{align}
Similarly, by employing eq. \eqref{eq:partialfhat},
\begin{equation}
\label{eq:cdadj} \hat{D}_{\mu}\longmapsto \hat{g}(\hat{x})\hat{D}_{\mu}\hat{g}^{-1}(\hat{x}) \ .
\end{equation}
Then, the noncommutative YM action can be written in two equivalent ways  \cite{Douglas:2001ba, Szabo:2001kg} 
\begin{align}
\label{eq:ncymop}
    S_{YM}=&-\frac{N}{2g^2}(2\pi)^{D/2}\mathrm{Pf}(\theta)\mathrm{Tr}\ \mathrm{tr}\left(\left[\hat{C}_{\mu},\hat{C}_{\nu}\right]+(\theta^{-1})_{\mu\nu}\right)^2 \nonumber \\
    =&-\frac{N}{2g^2}(2\pi)^{D/2}\mathrm{Pf}(\theta)\mathrm{Tr} \ \mathrm{tr} \left( \left[\hat{D}_{\mu},\hat{D}_{\nu}\right]^2 \right) \ .
\end{align}

\section{Gauge-invariant observables}
\label{sec:obs}
Without loss of generality we consider paths $C_v: \tau \in [0,1] \longmapsto  \xi^{\mu}(\tau)$ that start at the origin and end at $v$, so that $ \xi(0)=0$ and $ \xi(1)=v$. We define noncommutative Wilson lines with $x\in \mathbb{R}^D$ as base point  \cite{Szabo:2001kg}
\begin{align}
\label{eq:wxc}
    w(x; C_{v})=&\mathrm{P}\ \mathrm{exp}_{*}\left(i\int_{C_v}A_{\mu}(x+ \xi)d\xi^{\mu}\right) \nonumber \\
    =&\sum_{n=0}^{\infty}\int_0^1d\tau_1\int_{\tau_1}^1d\tau_2\dots \int_{\tau_{n-1}}^1d\tau_n\ \dot{ \xi}^{\mu_1}(\tau_1)\dot{ \xi}^{\mu_2}(\tau_2)\dots \dot{ \xi}^{\mu_n}(\tau_n)\times \nonumber \\
    &\times A_{\mu_1}(x+ \xi(\tau_1))* A_{\mu_2}(x+ \xi(\tau_2))*\dots * A_{\mu_n}(x+ \xi(\tau_n)) \ ,
\end{align}
which transform as
\begin{equation} \label{line000}
    w(x; C_{v})\longmapsto g(x)* w(x; C_{v}) * g^{-1}(x+v) 
\end{equation}
under gauge transformations. Now, given a local operator $O(x)$ transforming in the adjoint representation of the gauge group, the momentum-space operator
\begin{equation}
\label{eq:gaugeinv}
    \widetilde{O}(k_v)=\frac{1}{N}\mathrm{tr}\int d^Dx\  O(x)* w(x; C_{v})* e^{ik^{\nu}_vx_{\nu}}
\end{equation}
is gauge invariant according to eqs. \eqref{line000} and \eqref{eq:translcoord}, provided that $v_{\mu}=k^{\nu}_v\theta_{\nu\mu}$. The above definition easily generalizes to multiple operator insertions. \par 
For closed contours with no operator insertion we define the noncommutative Wilson loops as
\begin{equation}
	\label{eq:Wcapital}
	W(C_0)=\frac{1}{NV_D}\mathrm{tr}\int d^Dx\ w(x; C_0) \, ,
\end{equation}
where the normalization factor has been chosen such that the trivial term in the expansion of the path-ordered exponential is equal to $1$. We will momentarily write the above observables in terms of $\hat{D}_{\mu}$ and $\hat{C}_{\mu}$ defined in eq. \eqref{eq:cd} in the operator representation.

\subsection{Operator representation: First definition}

The first realization \cite{Szabo:2001kg} of the noncommutative Wilson line in the operator representation is given by 
\begin{equation}
\label{eq:w1}
    \mathrm{P}\ \mathrm{exp}\left(\int_{C_{v}} \hat{D}_{\mu}\ d\xi^{\mu}\right) \ .
\end{equation}
Though $\hat{D}_{\mu}$ in the line integral is path independent, the corresponding path-ordered exponential is far from being trivial: Since the components of $\hat{D}_{\mu}$ do not commute in general, it is not possible to pull the covariant derivative outside the contour integral. Yet, thanks to eq. \eqref{eq:cdadj}, the quantity in eq. \eqref{eq:w1} transforms in the adjoint representation of the gauge group
\begin{equation}
\label{eq:w1trans}
     \mathrm{P}\ \mathrm{exp}\left(\int_{C_{v}} \hat{D}_{\mu}\ d\xi^{\mu}\right)\longmapsto \hat{g}(\hat{x}) \, \mathrm{P}\ \mathrm{exp}\left(\int_{C_{v}} \hat{D}_{\mu}\ d\xi^{\mu}\right)\hat{g}^{-1}(\hat{x}) \ .
\end{equation}
We now express this operator in terms of a path-dependent connection. Let us first consider the path-ordered exponential
\begin{align}
	\label{eq:pexpa}
	\mathrm{P}\ \mathrm{exp}\left(i\int_{C_{v}} \hat{A}_{\mu}(\hat{x}+\xi)d\xi^{\mu}\right) \ .
\end{align}
We act on $\hat{A}_{\mu}(\hat{x}+\xi)$ by a gauge transformation living on the contour $C_v$ implemented by
\begin{equation}
\label{eq:uxi}
    \hat{U}(\xi(\tau))=e^{-\xi_{\mu}(\tau)\hat{\partial}^{\mu}} \ ,
\end{equation}
where $\tau\in[0,1]$ parametrizes the contour. We get
\begin{align}
\label{eq:formalgauge}
    &\hat{U}(0)\, \mathrm{P}\ \mathrm{exp}\left(i\int_{C_{v}} \hat{A}_{\mu}(\hat{x}+\xi)d\xi^{\mu}\right)\hat{U}^{-1}(v) \nonumber \\
    =&\mathrm{P}\ \mathrm{exp}\left(i\int_{C_{v}} \hat{U}(\xi)\hat{A}_{\mu}(\hat{x}+\xi)\hat{U}^{-1}(\xi)-i\hat{U}(\xi)\partial_{\mu}\hat{U}^{-1}(\xi) \ d\xi^{\mu}\right) \nonumber \\
    =& \mathrm{P}\ \mathrm{exp}\left(\int_{C_{v}} \hat{D}_{\mu}\ d\xi^{\mu}\right) \ ,
\end{align}
where we have employed eq. \eqref{eq:transl1}. Thus, we finally write
\begin{equation}
\label{eq:drop}
     \mathrm{P}\ \mathrm{exp}\left(\int_{C_{v}} \hat{D}_{\mu}\ d\xi^{\mu}\right)=\mathrm{P}\ \mathrm{exp}\left(i\int_{C_{v}} \hat{A}_{\mu}(\hat{x}+\xi)d\xi^{\mu}\right)e^{v_{\mu} \hat{\partial}^{\mu}}
\end{equation}
After these manipulations, it becomes evident how to express eq. \eqref{eq:gaugeinv} in terms of eq. \eqref{eq:w1}. Given a local operator $\hat{O}$ in the adjoint representation of the gauge group, we have
\begin{align}
    \widetilde{{O}}(k_v)=&\frac{1}{N}(2\pi)^{D/2}\mathrm{Pf}(\theta)\mathrm{Tr}\ \mathrm{tr} \left[\mathrm{P}\ \mathrm{exp}\left(\int_{C_{v}} \hat{D}_{\mu}\ d\xi^{\mu}\right) e^{-v^{\mu}\hat{\partial}_{\mu}}e^{ik_{v}^{\mu}\hat{x}_{\mu}}\hat{O}\right] \nonumber \\
    =& \frac{1}{N}\mathrm{tr}\int d^Dx\  O(x)* \mathrm{P}\ \mathrm{exp}_{*}\left(i\int_{C_v}A_{\mu}(x+ \xi)d\xi^{\mu}\right)* e^{ik^{\nu}_vx_{\nu}} \ ,
\end{align}
according to eq. \eqref{eq:gaugeinv} in the coordinate representation. In the special case of a Wilson loop along a closed path $C_0$, with the origin as base point and no operator insertion, we obtain
\begin{align}
	\label{eq:dictionary}
    W(C_0)=&\frac{1}{N\mathrm{Tr}(\hat{1})}\mathrm{Tr}\ \mathrm{tr} \ \mathrm{P}\ \mathrm{exp}\left(\oint_{C_{0}} \hat{D}_{\mu}\ d\xi^{\mu}\right) \nonumber \\
    =&\frac{1}{NV_D}\int d^Dx\ \mathrm{tr}\ \mathrm{P}\ \mathrm{exp}_{*}\left(i\oint_{C_0}A_{\mu}(x+ \xi)d\xi^{\mu}\right) \ ,
\end{align}
since $v=0$, where the normalization factor in the above equation has been chosen to match eq. \eqref{eq:Wcapital} in the coordinate representation.

\subsection{Operator representation: Second definition}

We introduce the object \cite{Gross:2000ba}
\begin{equation}
    \mathrm{P}\ \mathrm{exp}\left(\int_{C_{v}} \hat{C}_{\mu} d\xi^{\mu}\right) \ .
\end{equation}
For the same reasons recalled in the previous subsection, this operator is nontrivial and satisfies all the usual identities of the path-ordered exponentials. In particular, its gauge transformation law is
\begin{equation}
    \mathrm{P}\ \mathrm{exp}\left(\int_{C_{v}} \hat{C}_{\mu} d\xi^{\mu}\right)\longmapsto  \hat{g}(\hat{x})\ \mathrm{P}\ \mathrm{exp}\left(\int_{C_{v}} \hat{C}_{\mu} d\xi^{\mu}\right)\hat{g}^{-1}(\hat{x}) \ .
\end{equation}
Acting on the connection with the gauge transformation living on the contour
\begin{equation}
\label{eq:vxi}
    \hat{V}(\xi(\tau))=e^{i\xi^{\mu}(\tau)(\theta^{-1})_{\mu\nu}\hat{x}^{\nu}}
\end{equation}
and passing to a path-dependent connection by the same chain of identities as previously, except for the fact that now we employ eq. \eqref{eq:transl} instead of eq. \eqref{eq:transl1}, we obtain
\begin{align}
\label{eq:formalgauge1}
    &\hat{V}(0) \, \mathrm{P}\ \mathrm{exp}\left(i\int_{C_{v}} \hat{A}_{\mu}(\hat{x}+\xi)d\xi^{\mu}\right)\hat{V}^{-1}(v) \nonumber \\
    =&\mathrm{P}\ \mathrm{exp}\left(i\int_{C_{v}} \hat{V}(\xi)\hat{A}_{\mu}(\hat{x}+\xi)\hat{V}^{-1}(\xi)-i\hat{V}(\xi)\partial_{\mu}\hat{V}^{-1}(\xi) \ d\xi^{\mu}\right) \ .
\end{align}
To compute the affine term in the gauge-transformed connection, we employ eq. \eqref{eq:transl}
\begin{align}
	-i\hat{V}(\xi)\partial_{\mu}\hat{V}^{-1}(\xi)=&-e^{i\xi^{\alpha}(\tau)(\theta^{-1})_{\alpha\beta}\hat{x}^{\beta}}(\theta^{-1})_{\mu\nu}\hat{x}^{\nu}e^{-i\xi^{\alpha}(\tau)(\theta^{-1})_{\alpha\beta}\hat{x}^{\beta}} \nonumber \\
	=&-(\theta^{-1})_{\mu\nu}\hat{x}^{\nu}+(\theta^{-1})_{\mu\nu}\xi^{\nu}(\tau) \ ,
\end{align}
where in the second line we have used eq. \eqref{eq:transl}. Then, we get
\begin{align}
	\label{eq:v0vv}
	&\hat{V}(0) \, \mathrm{P}\ \mathrm{exp}\left(i\int_{C_{v}} \hat{A}_{\mu}(\hat{x}+\xi)d\xi^{\mu}\right)\hat{V}^{-1}(v) \nonumber \\
	=&\mathrm{P}\ \mathrm{exp}\left(\int_{C_{v}} \hat{C}_{\mu} d\xi^{\mu}\right) \mathrm{exp}\left(i\int_0^1d\tau\ \dot{\xi}^{\mu}(\tau)(\theta^{-1})_{\mu\nu}\xi^{\nu}(\tau)\right) 
\end{align}
and, finally,
\begin{align}
    \mathrm{P}\ \mathrm{exp}\left(\int_{C_{v}} \hat{C}_{\mu} d\xi^{\mu}\right)
    = \mathrm{P}\ \mathrm{exp}\left(i\int_{C_{v}} \hat{A}_{\mu}(\hat{x}+\xi)d\xi^{\mu}\right) e^{-iv_{\mu}(\theta^{-1})^{\mu\nu}\hat{x}_{\nu}}e^{-i\int_0^1d\tau\ \dot{\xi}_{\mu}(\tau)(\theta^{-1})^{\mu\nu}\xi_{\nu}(\tau)} \ ,
\end{align}
with a new central factor appearing in the final expression for the Wilson-line operator. \par
As a results, given a local operator $\hat{O}$ in the adjoint representation of the gauge group, we obtain the desired result
\begin{align}
    \widetilde{{O}}(k)=&\frac{1}{N}(2\pi)^{D/2}\mathrm{Pf}(\theta)\mathrm{Tr}\ \mathrm{tr} \left[\mathrm{P}\ \mathrm{exp}\left(\int_{C_{v}} \hat{C}_{\mu}\ d\xi^{\mu}\right)\hat{O}\right]e^{i\int_0^1d\tau\ \dot{\xi}_{\mu}(\tau)(\theta^{-1})^{\mu\nu}\xi_{\nu}(\tau)} \nonumber \\
    =& \frac{1}{N}\mathrm{tr}\int d^Dx\  O(x)* \mathrm{P}\ \mathrm{exp}_{*}\left(i\int_{C_v}A_{\mu}(x+ \xi)d\xi^{\mu}\right)* e^{ik^{\nu}_vx_{\nu}} \ ,
\end{align}
according to eq. \eqref{eq:gaugeinv} in the coordinate representation. In the special case of a Wilson loop along a closed path $C_0$ with the origin as base point and no operator insertion we get
\begin{align}
\label{eq:loopc}
    W(C_0)=&\frac{1}{N\mathrm{Tr}(\hat{1})}\mathrm{Tr}\ \mathrm{tr} \ \mathrm{P}\ \mathrm{exp}\left(\oint_{C_{0}} \hat{C}_{\mu}\ d\xi^{\mu}\right)\ e^{i\int_0^1d\tau\ \dot{\xi}_{\mu}(\tau)(\theta^{-1})^{\mu\nu}\xi_{\nu}(\tau)} \nonumber \\
    =&\frac{1}{NV_D}\int d^Dx\ \mathrm{tr}\ \mathrm{P}\ \mathrm{exp}_{*}\left(i\oint_{C_0}A_{\mu}(x+ \xi)d\xi^{\mu}\right) \ ,
\end{align}
since $v=0$, where the normalization factor in the above equation has been chosen to match eq. \eqref{eq:Wcapital} in the coordinate representation.

\section{V.e.v. of noncommutative Wilson loops in the planar limit}
\label{6}

We compute the v.e.v. of noncommutative Wilson loops in U($N$) YM theory to the leading large-$N$ order -- which by a standard argument (appendix \ref{app:ncfields}) coincides with the planar limit of the noncommutative theory -- in terms of the corresponding commutative objects. 
We start with the noncommutative Wilson loop
\begin{equation}
    W(C_0)=\frac{1}{NV_D}\int d^Dx\ \mathrm{tr}\ \mathrm{P}\ \mathrm{exp}_*\left(i\oint_{C_0}A_{\mu}(x+\xi)d\xi^{\mu}\right) \ ,
\end{equation}
where $C_0$ is a simple closed curve based at the origin, with no cusps or self-intersections.\par
We take its v.e.v. in the planar limit, denoting the planar correlators at finite $\theta_{\mu\nu}$ as $\expval{\dots }^{(\theta)}_{\mathrm{pl}}$ and the ones at $\theta_{\mu\nu}=0$ simply as $\expval{\dots }_{\mathrm{pl}}$. Connected correlators are denoted by adding a subscript $_{\rm conn}$, e.g. $\expval{\dots }^{(\theta)}_{\text{conn, pl}}$. 
Expanding the path-ordered exponential, we obtain 
\begin{align}
    \label{eq:expval}
    \expval{W(C_0)}_{\text{pl}}^{(\theta)}=
    &\frac{1}{N}\sum_{n=0}^{\infty}\int_0^1d\tau_1\dots \int_0^{\tau_{n-1}}d\tau_n\ \dot{\xi}^{\mu_1}(\tau_1)\dots \dot{\xi}^{\mu_n}(\tau_n)\mathrm{tr}\left(t^{a_1}\dots t^{a_n}\right) \nonumber \\
     & \expval{A_{\mu_1}^{a_1}(\xi(\tau_1))*\dots *A_{\mu_1}^{a_n}(\xi(\tau_n))}^{(\theta)}_{\text{pl}} \ ,
\end{align}
where, thanks to translation invariance, the $D$-dimensional spacetime volume $V_D$ has been factorized out. We explicitly write the Groenewold-Moyal product in the contour integrals above
\begin{equation}
    \expval{A_{\mu_1}^{a_1}(\xi(\tau_1))*\dots *A_{\mu_1}^{a_n}(\xi(\tau_n))}^{(\theta)}_{\text{pl}}=e^{\frac{i}{2}\sum\limits_{j<j'}\partial_j\wedge\partial_{j'}}\expval{A_{\mu_1}^{a_1}(\xi(\tau_1))\dots A_{\mu_1}^{a_n}(\xi(\tau_n))}^{(\theta)}_{\text{pl}} \ .
\end{equation}
The further $\theta$ dependence can be revealed by decomposing the correlators on the r.h.s. in connected correlators, and then employing eq. \eqref{eq:filk}
\begin{equation}
\label{eq:phase1}
    \expval{A_{\mu_1}^{a_1}(\xi(\tau_1))*\dots *A_{\mu_1}^{a_n}(\xi(\tau_n))}^{(\theta)}_{\text{pl}}=e^{\frac{i}{2}\sum\limits_{j<j'}\partial_j\wedge\partial_{j'}}\sum_{s\in \mathcal{P}_n}\prod_{s_i\in s}\expval{\prod_{k\in s_i}A_{\mu_k}^{a_k}(\xi(\tau_k))}^{(\theta)}_{\text{conn, pl}} \ ,
\end{equation}
where $\mathcal{P}_n$ denotes the set of partitions of $ \{1,2,\dots ,n\}$, $s=\{s_1,\dots, s_I\}$ is a given partition, $s_i$ is an element of the partition, $k$ are the integers belonging to $s_i$.\par
For each connected correlator in the r.h.s., applying eq. \eqref{eq:filk}, we get
\begin{equation}
\label{eq:phase2}
    \expval{\prod_{k\in s_i}A_{\mu_k}^{a_k}(\xi(\tau_k))}^{(\theta)}_{\text{conn, pl}}=\mathrm{exp}\left(\frac{i}{2}\sum_{\substack{\ell<\ell' \\
    \ell,\ell'\in s_i}}\partial_\ell\wedge \partial_{\ell'}\right)\expval{\prod_{k\in s_i}A_{\mu_k}^{a_k}(\xi(\tau_k))}_{\text{conn, pl}} \ .
\end{equation}
Then, we find
\begin{align}
\label{eq:phase3}
    \expval{A_{\mu_1}^{a_1}(\xi(\tau_1))*\dots *A_{\mu_1}^{a_n}(\xi(\tau_n))}^{(\theta)}_{\text{pl}}=&\sum_{s\in\mathcal{P}_n}\mathrm{exp}\left(\frac{i}{2}\sum_{j<j'}\partial_j\wedge \partial_{j'}\right) \nonumber \\
    & \prod_{s_i\in s}\mathrm{exp}\left(\frac{i}{2}\sum_{\substack{\ell<\ell' \\
    \ell,\ell'\in s_i}}\partial_\ell\wedge \partial_{\ell'}\right)\expval{\prod_{k\in s_i}A_{\mu_k}^{a_k}(\xi(\tau_k))}_{\text{conn, pl}} \ .
\end{align}
Substituting into eq. \eqref{eq:expval}, we finally obtain
\begin{align}
\label{eq:wlvev}
   & \expval{W(C_0)}_{\text{pl}}^{(\theta)} \nonumber \\
   &=\frac{1}{N}\sum_{n=0}^{\infty}\int_0^1d\tau_1\dots \int_0^{\tau_{n-1}}d\tau_n\ \dot{\xi}^{\mu_1}(\tau_1)\dots \dot{\xi}^{\mu_n}(\tau_n)\mathrm{tr}\left(t^{a_1}\dots t^{a_n}\right) \nonumber \\
  & \,\,\,\,\,\, \mathrm{exp}\left(\frac{i}{2}\sum_{j<j'}\partial_j\wedge \partial_{j'}\right)\sum_{s \in \mathcal{P}_n} \prod_{s_i\in s}\mathrm{exp}\left(\frac{i}{2}\sum_{\substack{\ell<\ell' \\
    \ell,\ell'\in s_i}}\partial_\ell\wedge \partial_{\ell'}\right)\expval{\prod_{k\in s_i}A_{\mu_k}^{a_k}(\xi(\tau_k))}_{\text{conn, pl}} \ ,
\end{align}
where now the whole $\theta$ dependence is inside the wedge products in the phase factors.

\section{Twistor Wilson loops}
\label{sec:twistwl}

\subsection{Definition}

In order to construct twistor Wilson loops, it is convenient to choose complex coordinates
\begin{equation}
\label{eq:coord}
    u=\frac{x^1+ix^2}{\sqrt{2}}\ , \qquad \bar{u}=\frac{x^1-ix^2}{\sqrt{2}}\ , \qquad  w=\frac{x^3+ix^4}{\sqrt{2}}\ , \qquad \bar{w}=\frac{x^3-ix^4}{\sqrt{2}} \ .
\end{equation}
for the noncommutative four-dimensional spacetime
\begin{align}
	[\hat{x}^{1},\hat{x}^{2}] = [\hat{x}^{3},\hat{x}^{4}]=i\vartheta \hat{1}\ .
\end{align}
with $\mathrm{Pf}(\theta)=\vartheta^2$. In the complex basis, the nonzero commutators of the coordinates are
\begin{align}
	\left[\hat{u}, \hat{\bar{u}}\right]=\left[\hat{w}, \hat{\bar{w}}\right]=\vartheta\hat{1} \ ,
\end{align}
with the metric tensor
\begin{align}
	\label{eq:gmunu}
	&g^{u\bar{u}}=g_{u\bar{u}}=1 \ , 
	&g^{w\bar{w}}=g_{w\bar{w}}=1\ 
\end{align}
and the nonzero elements of the noncommutativity matrix
\begin{align}
\label{eq:thetamunu}
	& \theta^{u\bar{u}}=-\theta^{\bar{u}u}=-\theta_{u\bar{u}}=\theta_{\bar{u}u}=-i\vartheta \nonumber \\
	& \theta^{w\bar{w}}=-\theta^{\bar{w}w}=-\theta_{w\bar{w}}=\theta_{\bar{w}w}=-i\vartheta \ .
\end{align}
We choose a closed path $C_{u\bar{u}}$, living on the plane $u\bar{u}$, with the origin as base point and without cusps and self-intersections
\begin{align}
    C_{u\bar{u}}:\  & [0,1] \longrightarrow \mathbb{R}^4 \nonumber \\
    & \tau \longmapsto \xi^{\mu}(\tau)=\left(\zeta(\tau),\bar{\zeta}(\tau),0,0\right)
\end{align}
and define the associated twistor Wilson loops \cite{Bochicchio:2012bj}
\begin{align}
	\label{TW1}
    &W_{\lambda}(C_{u\bar{u}})= \nonumber \\ 
    &\frac{1}{N\mathrm{Tr}'(\hat{1})}  \mathrm{Tr}'\int \frac{d^2u}{V_2}\ \mathrm{tr}\ \mathrm{P}\ \mathrm{exp}_{*'}\Big[i\oint_{C_{u\bar{u}}}\left(\hat{A}_u(u+\zeta,\bar{u}+\bar{\zeta},\hat{w},\hat{\bar{w}})+\lambda \hat{D}_w(u+\zeta,\bar{u}+\bar{\zeta},\hat{w},\hat{\bar{w}})\right)d\zeta \nonumber \\
    +&\left(\hat{A}_{\bar{u}}(u+\zeta,\bar{u}+\bar{\zeta},\hat{w},\hat{\bar{w}})+\lambda^{-1} \hat{D}_{\bar{w}}(u+\zeta,\bar{u}+\bar{\zeta},\hat{w},\hat{\bar{w}})\right)d\bar{\zeta}\Big] \ ,
\end{align}
where $V_2$ is the volume of two-dimensional spacetime, $\mathrm{Tr}'$ is the trace over the Fock space on which the noncommutative coordinates $\hat{w},\hat{\bar{w}}$ are represented, $\lambda\in\mathbb{C}/\{0\}$ and $*'$ denotes the Groenewold-Moyal product restricted to the coordinates $u,\bar{u}$, which is defined as
\begin{align}
    f_1(u_1,\bar{u}_1,\hat{w},\hat{\bar{w}})*'.&..*'f_n(u_n,\bar{u}_n,\hat{w},\hat{\bar{w}})= \nonumber \\
    =&\prod_{j<k}^n\mathrm{exp}\left(\frac{\vartheta}{2}\left(\frac{\partial}{\partial u_j}\frac{\partial}{\partial \bar{u}_k}-\frac{\partial}{\partial \bar{u}_j}\frac{\partial}{\partial u_k}\right)\right)\hat{f}_1(u_1,\bar{u}_1,\hat{w},\hat{\bar{w}})\dots \hat{f}_n(u_n,\bar{u}_n,\hat{w},\hat{\bar{w}}) \ .
\end{align}
Since the operator-valued covariant derivatives in eq. \eqref{TW1} transform in the adjoint representation of the gauge group, the twistor Wilson loops in eq. \eqref{TW1} are manifestly gauge invariant.\par
The covariant derivatives occurring in twistor Wilson loops of noncommutative pure YM theory play the same role as the scalar fields in SUSY Wilson loops of commutative theories with extended supersymmetry, including the factor of $i$ in front of the connections in the covariant derivatives  $\hat{D}_{w}= \hat{\partial}_w+i\hat{A}_w, \,\hat{D}_{\bar{w}}= \hat{\partial}_{\bar{w}}+i\hat{A}_{\bar{w}}$ in eq. \eqref{TW1}, in analogy with the factor of $i$ in front of the scalar fields in eq. \eqref{eq:susywilsonloop}.\par
Ultimately, as we will demonstrate in the next section, it is precisely the occurrence of the above factor of $i$, in combination with the Euclidean invariance of the commutative theory together with the induced residual rotational invariance of the planar noncommutative theory due to $\vartheta_1=\vartheta_2=\vartheta$, that is responsible for the triviality of the v.e.v. of twistor Wilson loops, somehow in analogy with the SUSY case.\par
Employing the identities in section \ref{sec:obs}, we express the twistor Wilson loops more concisely in the operator representation
\begin{equation}
\label{eq:twistorwl}
    W_{\lambda}(C_{u\bar{u}})=\frac{1}{N\mathrm{Tr}(\hat{1})} \mathrm{Tr}\ \mathrm{tr}\ \mathrm{P}\ \mathrm{exp}\left(\oint_{C_{u\bar{u}}}(\hat{D}_u+i\lambda\hat{D}_w)d\zeta+(\hat{D}_{\bar{u}}+i\lambda^{-1}\hat{D}_{\bar{w}})d\bar{\zeta}\right) \ .
\end{equation}
To understand the meaning of the above equation, it is convenient to write it into a new form. We introduce the closed paths $\mathcal{C}^{\lambda}$ living in the \emph{complexified} four-dimensional spacetime
\begin{align}
\label{eq:clambda}
    \mathcal{C}^{\lambda}: \ & [0,1]\longmapsto \mathbb{C}^4 \nonumber \\
    & \tau \longmapsto \zeta^{\mu}(\tau)=\left(\zeta(\tau),\bar{\zeta}(\tau),i\lambda \zeta(\tau), i\lambda^{-1}\bar{\zeta}(\tau)\right)
\end{align}
These paths and their velocities are supported on the submanifold
\begin{align}
	&(u,\bar{u}, w, \bar{w})=(u,\bar{u}, i\lambda u, i\lambda^{-1}\bar{u}) \nonumber \\
	&(\dot{u},\dot{\bar{u}}, \dot{w}, \dot{\bar{w}})=(\dot{u},\dot{\bar{u}}, i\lambda \dot{u}, i\lambda^{-1}\dot{\bar{u}})
\end{align}
that is Lagrangian \cite{Bochicchio:2012bj} with respect to the K\"ahler form $du\wedge d\bar{u}+dw\wedge d\bar{w}$ and satisfies
\begin{equation}
\label{eq:zetazeta}
    \zeta^{\mu}(\tau)\zeta_{\mu}(\tau')=\dot{\zeta}^{\mu}(\tau)\zeta_{\mu}(\tau')=\dot{\zeta}^{\mu}(\tau)\dot{\zeta}_{\mu}(\tau')=0\ , \qquad \forall\  \tau,\tau'\in [0,1]\ ,\  \lambda \in \mathbb{C}/\{0\} \, .
\end{equation}
In terms of the complexified paths, eq. \eqref{eq:twistorwl} can be written more compactly as
\begin{equation}
	\label{eq:wlambda}
    W_{\lambda}(C_{u\bar{u}})=\frac{1}{N\mathrm{Tr}(\hat{1})} \mathrm{Tr}\ \mathrm{tr}\ \mathrm{P}\ \mathrm{exp}\left(\oint_{\mathcal{C}^{\lambda}}\hat{D}_{\mu}d\zeta^{\mu} \right) 
\end{equation}
that, according to eq. \eqref{eq:dictionary}, is equivalent to
\begin{align}
	\label{eq:twistorcoord}
	W_{\lambda}(C_{u\bar{u}})=&\frac{1}{N\mathrm{Tr}(\hat{1})}\mathrm{tr}\int d^4x\ \mathrm{P}\ \mathrm{exp}\left(i\oint_{\mathcal{C}^{\lambda}}\hat{A}_{\mu}(\hat{x}+\zeta)d\zeta^{\mu}\right) \nonumber \\
	=&\frac{1}{NV_4}\mathrm{tr}\int d^4x\ \mathrm{P}\ \mathrm{exp}_*\left(i\oint_{\mathcal{C}^{\lambda}}A_{\mu}(x+\zeta)d\zeta^{\mu}\right)
\end{align}
as we may verify by employing the gauge transformation
\begin{align}
    \hat{S}(\xi(\tau))=& \mathrm{exp}\left(-\zeta(\tau)\hat{\partial}_u-\bar{\zeta}(\tau)\hat{\partial}_{\bar{u}}-i\lambda \zeta(\tau)\hat{\partial}_{w}-i\lambda^{-1}\bar{\zeta}(\tau)\hat{\partial}_{\bar{w}}\right) \nonumber \\
    =& \mathrm{exp}\left(-\zeta^{\mu}(\tau)\hat{\partial}_{\mu}\right)
\end{align}
that, contrary to the $\hat{U}$ and $\hat{V}$ in eqs. \eqref{eq:uxi} and \eqref{eq:vxi} respectively, is nonunitary. Yet, though this gauge transformation is not a symmetry of the noncommutative action, it leaves invariant the twistor Wilson loops because of the trace involved in their definition.\par
\par Alternatively, from the twistor Wilson loops in terms of the $\hat{C}_{\mu}$
\begin{equation}
	\label{eq:twistorwlC}
	W_{\lambda}(C_{u\bar{u}})=\frac{1}{N\mathrm{Tr}(\hat{1})} \mathrm{Tr}\ \mathrm{tr}\ \mathrm{P}\ \mathrm{exp}\left(\oint_{C_{u\bar{u}}}(\hat{C}_u+i\lambda\hat{C}_w)d\zeta+(\hat{C}_{\bar{u}}+i\lambda^{-1}\hat{C}_{\bar{w}})d\bar{\zeta}\right) \ ,
\end{equation}
by the gauge transformation
\begin{align}
	\hat{S}'(\xi(\tau))=& \mathrm{exp}\left(\zeta(\tau)\vartheta^{-1}\hat{\bar{u}}-\bar{\zeta}(\tau)\vartheta^{-1}\hat{u}+i\lambda \zeta(\tau)\vartheta^{-1}\hat{\bar{w}}-i\lambda^{-1}\bar{\zeta}(\tau)\vartheta^{-1}\hat{w}\right)
\end{align}
analogous to eq. \eqref{eq:v0vv}, we obtain as expected
\begin{align}
	\label{eq:wlambdaC}
	W_{\lambda}(C_{u\bar{u}})=&\frac{1}{N\mathrm{Tr}(\hat{1})} \mathrm{Tr}\ \mathrm{tr}\  \mathrm{P}\ \mathrm{exp}\left(\oint_{\mathcal{C}^{\lambda}}\hat{A}_{\mu}(\hat{x}+\zeta)d\zeta^{\mu} \right) \nonumber \\
	=& \frac{1}{NV_4}\mathrm{tr}\int d^4x\ \mathrm{P}\ \mathrm{exp}_*\left(i\oint_{\mathcal{C}^{\lambda}}A_{\mu}(x+\zeta)d\zeta^{\mu}\right) \ ,
\end{align}
the phase factor in the second line of eq. \eqref{eq:v0vv} being trivial because, for the $\theta^{\mu\nu}$ in eq. \eqref{eq:thetamunu}, the second line of eq. \eqref{eq:zetazetadot} implies
\begin{align}
	\dot{\zeta}^{\mu}(\tau)(\theta^{-1})_{\mu\nu}\zeta^{\nu}(\tau)=0  \ .
\end{align}

\subsection{Triviality of the planar v.e.v. of twistor Wilson loops}
\label{sec:triviality}

We now prove that, to all orders of perturbation theory and to the leading large-$N$ order, in noncommutative U($N$) YM theory the v.e.v. of twistor Wilson loops in eq. \eqref{eq:twistorwl} is trivial -- actually, to all orders in the $\vartheta$ expansion. 
\par We employ the second line of eq. \eqref{eq:twistorcoord} and the corresponding computation in eq. \eqref{eq:wlvev} to get
\begin{align}
\label{eq:wlvevzeta}
 &\expval{W_{\lambda}(C_{u\bar{u}})}_{\text{pl}}^{(\theta)} \nonumber \\
    &=\frac{1}{N}\sum_{n=0}^{\infty}\int_0^1d\tau_1\dots \int_0^{\tau_{n-1}}d\tau_n\ \dot{\zeta}^{\mu_1}(\tau_1)\dots \dot{\zeta}^{\mu_n}(\tau_n)\mathrm{tr}\left(t^{a_1}\dots t^{a_n}\right) \nonumber \\
    & \,\,\,\,\,\, \mathrm{exp}\left(\frac{i}{2}\sum_{j<j'}\partial_j\wedge \partial_{j'}\right)\sum_{s \in \mathcal{P}_n} \prod_{s_i\in s}\mathrm{exp}\left(\frac{i}{2}\sum_{\substack{\ell<\ell' \\
    \ell,\ell'\in s_i}}\partial_\ell\wedge \partial_{\ell'}\right)\expval{\prod_{k\in s_i}A_{\mu_k}^{a_k}(\zeta(\tau_k))}_{\text{conn, pl}} \, ,
\end{align}
where an intermediate regularization of the above correlators -- discussed at the end of this section -- is understood. Now, Euclidean translational and rotational invariance imply that the commutative planar correlators (in the Feynman gauge, which is Euclidean invariant)
\label{eq:integrand}
\begin{equation}
\label{eq:corr}
    \expval{\prod_{i\in s}A_{\mu_i}^{a_i}(\zeta(\tau_i))}_{\text{conn, pl}} \, ,
\end{equation}
to a given order of perturbation theory, consist of the product of polynomials in the metric $g_{\mu\nu}$ in eq. \eqref{eq:gmunu} and the differences $\left(\zeta^{\mu}(\tau_i)-\zeta^{\mu}(\tau_j)\right)$ multiplied by scalar functions of the differences. \par
Using the definition of the paths in eq. \eqref{eq:clambda}, we immediately see that the zero-order term in $\theta^{\mu\nu}$ in eq. \eqref{eq:wlvevzeta} vanishes because, by Euclidean covariance, the only possible contractions are between the velocities $\dot\zeta^{\mu}(\tau_i)$ in the second line of eq. \eqref{eq:wlvevzeta} and the aforementioned polynomials in $g_{\mu\nu}$ and $\left(\zeta^{\mu}(\tau_j)-\zeta^{\mu}(\tau_k)\right)$ arising from the commutative correlators in eq. \eqref{eq:corr} 
\begin{align}
	\label{eq:zetacontractions}
    \dot\zeta^{\nu}(\tau_i)\dot\zeta_{\nu}(\tau_j)=&\dot\zeta(\tau_i)\dot{\bar{\zeta}}(\tau_j)+\dot{\bar{\zeta}}(\tau_i)\dot\zeta(\tau_j)+i\lambda\dot\zeta(\tau_i)i\lambda^{-1}\dot{\bar{\zeta}}(\tau_j)+i\lambda\dot{\bar{\zeta}}(\tau_i)i\lambda^{-1}\dot\zeta(\tau_j) \nonumber \\
    =&\dot\zeta(\tau_i)\dot{\bar{\zeta}}(\tau_j)+\dot{\bar{\zeta}}(\tau_i)\dot\zeta(\tau_j)-\dot\zeta(\tau_i)\dot{\bar{\zeta}}(\tau_j)-\dot{\bar{\zeta}}(\tau_i)\dot\zeta(\tau_j)=0 \nonumber \\
    \dot\zeta^{\nu}(\tau_i)\zeta_{\nu}(\tau_j)=&\dot\zeta(\tau_i)\bar{\zeta}(\tau_j)+\dot{\bar{\zeta}}(\tau_i)\zeta(\tau_j)+i\lambda\dot\zeta(\tau_i)i\lambda^{-1}\bar{\zeta}(\tau_j)+i\lambda\dot{\bar{\zeta}}(\tau_i)i\lambda^{-1}\zeta(\tau_j) \nonumber \\
    =&\dot\zeta(\tau_i)\bar{\zeta}(\tau_j)+\dot{\bar{\zeta}}(\tau_i)\zeta(\tau_j)-\dot\zeta(\tau_i)\bar{\zeta}(\tau_j)-\dot{\bar{\zeta}}(\tau_i)\zeta(\tau_j)=0\ .
\end{align}
The same statement holds for the higher-order terms in $\theta^{\mu\nu}$. In this case, there are additional contributions arising from the derivatives in the phase factors in the third line of eq. \eqref{eq:wlvevzeta} that are proportional to the noncommutativity matrix $\theta^{\mu\nu}$. The derivatives act both on the aforementioned polynomials in $g_{\mu\nu}$ and the differences $\left(\zeta^{\mu}(\tau_i)-\zeta^{\mu}(\tau_j)\right)$ and on the scalar factors. Under the actions of the derivatives the following structures may be produced
\begin{align}
	\label{eq:contractions}
    &\zeta^{\mu}(\tau_i){\theta_{\mu}}^{\mu_1}{\theta_{\mu_1}}^{\mu_2}\dots {\theta_{\mu_{k-1}}}^{\mu_{k}}\theta_{\mu_{k}\nu}\zeta^{\nu}(\tau_j) \nonumber \\
    &\dot{\zeta}^{\mu}(\tau_i){\theta_{\mu}}^{\mu_1}{\theta_{\mu_1}}^{\mu_2}\dots {\theta_{\mu_{k-1}}}^{\mu_{k}}\theta_{\mu_{k}\nu}\zeta^{\nu}(\tau_j) \nonumber \\
    &\dot{\zeta}^{\mu}(\tau_i){\theta_{\mu}}^{\mu_1}{\theta_{\mu_1}}^{\mu_2}\dots {\theta_{\mu_{k-1}}}^{\mu_{k}}\theta_{\mu_{k}\nu}\dot{\zeta}^{\nu}(\tau_j) \ ,
\end{align}
where for $k=0$ the product of $\theta$'s should be read as just $\theta_{\mu\nu}$. Moreover, additional structures involving scalar combinations of $\theta$'s
\begin{equation}
	\label{eq:thetatrace}
	{\theta_{\nu}}^{\mu_1}{\theta_{\mu_1}}^{\mu_2}\dots {\theta_{\mu_{k-1}}}^{\mu_{k}}{\theta_{\mu_{k}}}^{\nu}
\end{equation}
may contribute to the noncommutative correlators.
\par Because of the tensor nature of the $\theta^{\mu\nu}$, eqs. \eqref{eq:contractions} and \eqref{eq:thetatrace} exhaust all the possible extra structures with respect to eq. \eqref{eq:zetacontractions}.
\par Now, each term in the expansion in powers of $\theta^{\mu\nu}$ of eq. \eqref{eq:wlvevzeta} necessarily contains a factor occurring in eq. \eqref{eq:contractions}, eventually multiplying a factor in eq. \eqref{eq:thetatrace}. Therefore, for the triviality of the v.e.v. in eq. \eqref{eq:wlvevzeta}, it suffices to demonstrate that all the structures in eq. \eqref{eq:contractions} vanish. To this aim, we proceed as follows. \par
We arrange the components of $g_{\mu\nu}$ and $\theta^{\mu\nu}$ in eqs. \eqref{eq:gmunu} and \eqref{eq:thetamunu} in two matrices
\begin{align}
	& g_{\mu\nu}=g^{\mu\nu}=\begin{pmatrix}
		0 & 1 & 0 & 0 \\
		1 & 0 & 0 & 0 \\
		0 & 0 & 0 & 1 \\
		0 & 0 & 1 & 0 \\
	\end{pmatrix} \ , && \theta^{\mu\nu}=-\theta_{\mu\nu}=\begin{pmatrix}
		0 & -i\vartheta & 0 & 0  \\
		+i\vartheta & 0 & 0 & 0  \\
		0 & 0 & 0 & -i\vartheta \\
		0 & 0 & +i\vartheta & 0  \\  
	\end{pmatrix} \ ,
\end{align}
where $\mu,\nu=w,\bar{w}, u, \bar{u}$. This implies
\begin{equation}
    {\theta^{\mu}}_{\nu}=-{\theta_{\mu}}^{\nu}=\begin{pmatrix}
        -i\vartheta & 0 & 0 & 0  \\
        0 & +i\vartheta & 0 & 0  \\
        0 & 0 & -i\vartheta & 0 \\
        0 & 0 & 0 & +i\vartheta  \\ 
    \end{pmatrix}
\end{equation}
and, hence,
\begin{equation}
    {\theta_{\mu}}^{\mu_1}{\theta_{\mu_1}}^{\mu_2}\dots {\theta_{\mu_{k-1}}}^{\mu_{k}}\theta_{\mu_{k}\nu}=\begin{pmatrix}
        0 & (i\vartheta)^{k+1} & 0 & 0  \\
       -(i\vartheta)^{k+1} & 0 & 0 & 0  \\
        0 & 0 & 0 & (i\vartheta)^{k+1} \\
        0 & 0 & -(i\vartheta)^{k+1} & 0 \\
    \end{pmatrix}
\end{equation}
It follows that
\begin{align}
	\label{eq:zetazetadot}
     \zeta^{\mu}(\tau_i){\theta_{\mu}}^{\mu_1}{\theta_{\mu_1}}^{\mu_2}&\dots {\theta_{\mu_{k-1}}}^{\mu_{k}}\theta_{\mu_{k}\nu}\zeta^{\nu}(\tau_j)= \nonumber \\
     =&(i\vartheta)^{k+1}\left(\zeta(\tau_i)\bar{\zeta}(\tau_j)+\bar{\zeta}(\tau_i)\zeta(\tau_j)-\zeta(\tau_i)\bar{\zeta}(\tau_j)-\bar{\zeta}(\tau_i)\zeta(\tau_j)\right)=0 \nonumber \\
     \dot{\zeta}^{\mu}(\tau_i){\theta_{\mu}}^{\mu_1}{\theta_{\mu_1}}^{\mu_2}&\dots {\theta_{\mu_{k-1}}}^{\mu_{k}}\theta_{\mu_{k}\nu}\zeta^{\nu}(\tau_j)= \nonumber \\
     =&(i\vartheta)^{k+1}\left(\dot{\zeta}(\tau_i)\bar{\zeta}(\tau_j)+\dot{\bar{\zeta}}(\tau_i)\zeta(\tau_j)-\dot{\zeta}(\tau_i)\bar{\zeta}(\tau_j)-\dot{\bar{\zeta}}(\tau_i)\zeta(\tau_j)\right)=0 \nonumber\\
     \dot{\zeta}^{\mu}(\tau_i){\theta_{\mu}}^{\mu_1}{\theta_{\mu_1}}^{\mu_2}&\dots {\theta_{\mu_{k-1}}}^{\mu_{k}}\theta_{\mu_{k}\nu}\dot{\zeta}^{\nu}(\tau_j)= \nonumber \\
     =&(i\vartheta)^{k+1}\left(\dot{\zeta}(\tau_i)\dot{\bar{\zeta}}(\tau_j)+\dot{\bar{\zeta}}(\tau_i)\dot{\zeta}(\tau_j)-\dot{\zeta}(\tau_i)\dot{\bar{\zeta}}(\tau_j)-\bar{\zeta}(\tau_i)\dot{\zeta}(\tau_j)\right)=0 \ .
\end{align}
Hence, we conclude that
\begin{equation}
\label{eq:trivial}
    \expval{W_{\lambda}(C_{u\bar{u}})}_{\text{pl}}^{(\theta)}=1
\end{equation}
regardless of the shape of $C_{u\bar{u}}$. \par
\par The above proof of triviality may be spoiled by singularities of the scalar contribution to the correlators in eq. \eqref{eq:corr}, which may arise from the vanishing of $(\zeta^{\nu}(\tau_i)-\zeta^{\nu}(\tau_j))(\zeta_{\nu}(\tau_k)-\zeta_{\nu}(\tau_\ell))$ for the very same reasons recalled above. To cure this problem, we need to choose a regularization -- preserving Euclidean invariance --  that makes the propagators (and, hence, the correlators) nonsingular at coinciding points. The simplest choice consists in deforming the propagators as \cite{Polchinski:1983gv}
\begin{align}
\label{eq:regulator}
     \frac{1}{4\pi^2 x^2} & \longrightarrow \frac{1-e^{-x^2/{2a^2}}}{4\pi^2x^2} & \text{(position space)} \ ,\nonumber \\
     \frac{1}{p^2} & \longrightarrow \frac{1}{p^2}e^{-p^2a^2}  & \text{(momentum space)} \ ,
\end{align}
where the regulator $a$ is a length scale. Since the regulated correlators with $n\geq 1 $ in eq. \eqref{eq:wlvevzeta} vanish provided that the regulator preserves -- as it does -- Euclidean invariance, we can remove the regulator without adding counterterms.

\subsection{Generalized twistor Wilson loops}

In noncommutative U($N$) YM theory, we can also define generalized twistor Wilson loops
\begin{equation}
    W_{\lambda,\mu}(C_{u\bar{u}})=\frac{1}{N\mathrm{Tr}(\hat{1})}\mathrm{Tr}\ \mathrm{tr}\ \mathrm{P}\ \mathrm{exp}\left(\oint_{C_{u\bar{u}}}(\mu\hat{D}_u+i\lambda\hat{D}_w)d\zeta+(\mu^{-1}\hat{D}_{\bar{u}}+i\lambda^{-1}\hat{D}_{\bar{w}})d\bar{\zeta}\right) \ ,
\end{equation}
where $\mu$ is a nonzero complex number. By an argument entirely analog to the one in the previous section, their v.e.v. to the leading large-$N$ order is trivial as well.

\section{Conclusions} \label{conclusion}

In the present paper we have demonstrated that, in the planar limit of both noncommutative and commutative YM theory, certain nontrivial Wilson loops exist that, at the leading large-$N$ order and to all orders of perturbation theory, have trivial v.e.v., somehow in analogy with certain SUSY Wilson loops in YM theories with extended supersymmetry. \par
As recalled in the introduction, the existence of such twistor Wilson loops is the starting point for identifying a conjecturally solvable  sector \cite{Bochicchio:2016toi} of large-$N$ YM theory, via the corresponding topological field/string theory trivial to the leading large-$N$ order, but nontrivially extended to the next-to-leading order in the large-$N$ expansion to include nonperturbative information on the glueballs \cite{Bochicchio:2016toi, Bochicchio2025}.
\newpage

\appendix
\section{Noncommutative spacetime}
\label{app:ncspaces}

\subsection{Representations}

By the Stone -- von Neumann theorem \cite{Szabo:2001kg}, the algebra in eq. \eqref{eq:ncalg} has a unique irreducible representation on a Hilbert space $\mathcal{H}$ up to unitary equivalence. To find it, we pass to the Darboux basis \cite{Szabo:2001kg}, where the commutation relations become
\begin{align}
     &\left[\hat{x}^{2\alpha-1},\hat{x}^{2\alpha}\right]=i\vartheta_{\alpha} && \alpha=1,2,\dots ,\frac{D}{2} \ .
\end{align}
Then, we further change basis to
\begin{equation}
    \hat{c}_{\alpha}=\frac{\hat{x}^{2\alpha-1}+i\ \mathrm{sgn}(\vartheta_{\alpha})\hat{x}^{2\alpha}}{\sqrt{2|\vartheta_{\alpha}|}}\ , \qquad \hat{c}_{\alpha}^{\dagger}=\frac{\hat{x}^{2\alpha-1}-i\ \mathrm{sgn}(\vartheta_{\alpha})\hat{x}^{2\alpha}}{\sqrt{2|\vartheta_{\alpha}|}} \ .
\end{equation}
These new operators satisfy the commutation relations of creation and annihilation operators
\begin{align}
     &\left[\hat{c}_{\alpha}, c_{\beta}\right]=\left[\hat{c}_{\alpha}^{\dagger}, c_{\beta}^{\dagger}\right]=0 && \left[\hat{c}_{\alpha}, c_{\beta}^{\dagger}\right]=\delta_{\alpha\beta}\hat{1}
\end{align}
and define the corresponding Fock representation
\begin{equation}
    \mathcal{H}=\overline{\bigoplus_{\vec{n}\in \mathbb{Z}_+^{D/2}}\mathbb{C}\ket{\vec{n}}} \ ,
\end{equation}
where the vectors $\ket{\vec{n}}$ read
\begin{align}
\label{eq:state}
	\ket{\vec{n}}=\prod_{\alpha=1}^{D/2}\frac{(\hat{c}_\alpha^{\dagger})^{n_i}}{\sqrt{n_\alpha!}}|\vec{0}\rangle \ ,
\end{align}
with $\vec{n}=(n_1,\dots, n_{D/2})$ and the vacuum $|\vec{0}\rangle$ annihilated by all the $\hat{c}_{\alpha}$. \par
The corresponding matrix elements of $\hat{c}_{\alpha}, \hat{c}_{\alpha}^{\dagger}$ read
\begin{equation}
    \hat{c}_{\alpha}\ket{\vec{n}}=\sqrt{n_{\alpha}}\ket{\vec{n}-\vec{1}_{\alpha}}\ , \qquad \hat{c}_{\alpha}^{\dagger}\ket{\vec{n}}=\sqrt{n_{\alpha}+1}\ket{\vec{n}+\vec{1}_{\alpha}}
\end{equation}
where $(\vec{1}_{\alpha})_{\beta}=\delta_{\alpha\beta}$.

\subsection{Trace}

\label{app:trace}

Given the definition of the trace of an operator $\hat{O}$ in the Fock representation $\mathcal{H}$
\begin{equation}
    \mathrm{Tr}\left[\hat{O}\right]=\sum_{\vec{n}\in\mathbb{Z}_+^{D/2}}\bra{\vec{n}}\hat{O}\ket{\vec{n}} \ ,
\end{equation}
we define the smaller subspaces
\begin{equation}
	\label{eq:nalpha}
    \mathcal{H}_\alpha=\overline{\bigoplus_{\substack{\vec{n}_{\alpha}\in \mathbb{Z}_+^{D/2} \\ (\vec{n}_\alpha)_{\beta}=n_{\alpha}\delta_{\alpha\beta}}}\mathbb{C}\ket{\vec{n}_\alpha}}\ , 
\end{equation}
According to the above decomposition, we get
\begin{equation}
	\label{eq:decomposition}
    \mathrm{Tr}\left[ e^{ik_{\mu}\hat{x}^{\mu}}\right]=\prod_{\alpha=1}^{D/2}\mathrm{Tr}_{\mathcal{H}_{\alpha}}\left[ e^{ik_{2\alpha-1}\hat{x}^{2\alpha-1}+ik_{2\alpha}\hat{x}^{2\alpha}}\right]
\end{equation}
We also assume without loss of generality that $\vartheta_\alpha>0$ for all $\alpha$. Thanks to the above factorization, we are free to separately consider each subspace
that simplifies proving eq. \eqref{eq:trace0} below as well, along the following lines. \par
Coherent states satisfy the completeness relation
\begin{equation}
\label{eq:product}
    \int \frac{d^2\beta_{\alpha}}{\pi}\ket{\beta_{\alpha}}\bra{\beta_{\alpha}}=\sum_{\substack{\vec{n}_{\alpha}\in \mathbb{Z}_+^{D/2} \\ (\vec{n}_\alpha)_{\beta}=n_{\alpha}\delta_{\alpha\beta}}}\ket{\vec{n}_{\alpha}}\bra{\vec{n}_{\alpha}}=\mathds{1}_{\mathcal{H}_{\alpha}} \ ,
\end{equation}
where $\vec{n}_{\alpha}$ has been defined in eq. \eqref{eq:nalpha} and
\begin{equation}
	\label{eq:betaalpha}
    \ket{\beta_\alpha}=\hat{U}_{\alpha}(\beta_{\alpha})\ket{\vec{0}}=e^{-|\beta_{\alpha}|^2/2}\sum_{n_{\alpha}=0}^{\infty} \frac{\beta^{n_{\alpha}}}{\sqrt{n_{\alpha}!}}\ket{\vec{n}_{\alpha}} \ ,
\end{equation}
with
\begin{equation}
\label{eq:ualpha}
    \hat{U}_{\alpha}(\beta_{\alpha})=\mathrm{exp}\left(\beta_{\alpha} \hat{c}_{\alpha}^*-\beta_{\alpha}^*\hat{c}_{\alpha}\right) \ .
\end{equation} 
Then, the trace over the subspace $\mathcal{H}_{\alpha}$ reads
\begin{align}
	\label{eq:tracecoherent}
	\mathrm{Tr}_{\mathcal{H}_{\alpha}}[\hat{O}]=&\sum_{\substack{\vec{n}_{\alpha}\in \mathbb{Z}_+^{D/2} \\ (\vec{n}_\alpha)_{\beta}=n_{\alpha}\delta_{\alpha\beta}}}\bra{\vec{n}_\alpha}\hat{O}\ket{\vec{n}_{\alpha}} \nonumber \\
	=&\int\frac{d^2\beta_{\alpha}}{\pi}\sum_{\substack{\vec{n}_{\alpha}\in \mathbb{Z}_+^{D/2} \\ (\vec{n}_\alpha)_{\beta}=n_{\alpha}\delta_{\alpha\beta}}}\bra{n_\alpha}\hat{O}\ket{\beta_{\alpha}}\bra{\beta_{\alpha}}\ket{\vec{n}_{\alpha}} \nonumber \\
	=& \int\frac{d^2\beta_{\alpha}}{\pi}\sum_{\substack{\vec{n}_{\alpha}\in \mathbb{Z}_+^{D/2} \\ (\vec{n}_\alpha)_{\beta}=n_{\alpha}\delta_{\alpha\beta}}}\bra{\beta_{\alpha}}\ket{\vec{n}_{\alpha}}\bra{n_\alpha}\hat{O}\ket{\beta_{\alpha}}=\int\frac{d^2\beta_{\alpha}}{\pi}\bra{\beta_{\alpha}}\hat{O}\ket{\beta_{\alpha}}
\end{align}
We write eq. \eqref{eq:decomposition} in terms of the $\hat{U}_{\alpha}$ operators
\begin{align}
	\mathrm{Tr}_{\mathcal{H}_{\alpha}}\left[ e^{ik_{2\alpha-1}\hat{x}^{2\alpha-1}+ik_{2\alpha}\hat{x}^{2\alpha}}\right]=\mathrm{Tr}_{\mathcal{H}_{\alpha}}\left[\hat{U}_{\alpha}(\gamma_{\alpha})\right] \ ,
\end{align}
with
\begin{equation}
	\label{eq:gammaalpha}
	\gamma_{\alpha}=\sqrt{\frac{\vartheta_{\alpha}}{2}}(ik_{2\alpha-1}-k_{2\alpha}) \ .
\end{equation}
Hence,
\begin{align}
    	\mathrm{Tr}_{\mathcal{H}_{\alpha}}\left[ e^{ik_{2\alpha-1}\hat{x}^{2\alpha-1}+ik_{2\alpha}\hat{x}^{2\alpha}}\right]=&\mathrm{Tr}_{\mathcal{H}_{\alpha}}\left[\hat{U}_{\alpha}(\gamma_{\alpha})\right] \nonumber \\
    =&\int\frac{d^2\beta_{\alpha}}{\pi}\bra{\beta_{\alpha}}\hat{U}_{\alpha}(\gamma_{\alpha})\ket{\beta_{\alpha}} \nonumber \\
    =&\int\frac{d^2\beta_{\alpha}}{\pi}\bra{\vec{0}}\hat{U}_{\alpha}(-\beta_{\alpha})\hat{U}_{\alpha}(\gamma_{\alpha})\hat{U}_{\alpha}(\beta_{\alpha})\ket{\vec{0}} \ ,
\end{align}
where in the second line above we have employed eq. \eqref{eq:tracecoherent}. From the Baker-Campbell-Hausdorff formula and eq. \eqref{eq:ncalg}, we obtain
\begin{equation}
    \hat{U}_{\alpha}(-\beta_{\alpha})\hat{U}_{\alpha}(\gamma_{\alpha})\hat{U}_{\alpha}(\beta_{\alpha})=\hat{U}_{\alpha}(\gamma_{\alpha})\ \mathrm{exp}(\beta_{\alpha}^*\gamma_{\alpha}-\gamma_{\alpha}^*\beta_{\alpha}) \ .
\end{equation}
Writing the phase factor as
\begin{equation}
    \mathrm{exp}(\beta_{\alpha}^*\gamma_{\alpha}-\gamma_{\alpha}^*\beta_{\alpha})=\mathrm{exp}\left(2i\Re(\beta_{\alpha})\Im(\gamma_{\alpha})-2i\Im(\beta_{\alpha})\Re(\gamma_{\alpha})\right)
\end{equation}
and the integration measure as
\begin{equation}
    \int \frac{d^2\beta_{\alpha}}{\pi}=\frac{1}{\pi}\int_{-\infty}^{+\infty}d\Re(\beta_{\alpha})\int_{-\infty}^{+\infty}d\Im(\beta_{\alpha}) \ ,
\end{equation}
we get
\begin{align}
    	\mathrm{Tr}_{\mathcal{H}_{\alpha}}\left[ e^{ik_{2\alpha-1}\hat{x}^{2\alpha-1}+ik_{2\alpha}\hat{x}^{2\alpha}}\right]=&\mathrm{Tr}_{\mathcal{H}_{\alpha}}\left[\hat{U}_{\alpha}(\gamma_{\alpha})\right] \nonumber \\
    	=&\frac{1}{\pi}(2\pi)^2 \delta(2\Im(\gamma_{\alpha}))\delta(2\Re(\gamma_{\alpha}))
\end{align}
Going back to the variables $k_{2\alpha-1}, k_{2\alpha}$ according to eq. \eqref{eq:gammaalpha}, we find
\begin{align}
	\delta(2\mathrm{Re}(\gamma_{\alpha}))\delta(2\mathrm{Im}(\gamma_{\alpha}))=&\delta\left(\sqrt{2\vartheta_{\alpha}}k_{2\alpha-1}\right)\delta\left(\sqrt{2\vartheta_{\alpha}}k_{2\alpha}\right) \nonumber \\
	=&\frac{1}{2\vartheta_{\alpha}}\delta(k_{2\alpha-1})\delta(k_{2\alpha})
\end{align}
and, hence,
\begin{equation}
    \mathrm{Tr}_{\mathcal{H}_{\alpha}}\left[ e^{ik_{2\alpha-1}\hat{x}^{2\alpha-1}+ik_{2\alpha}\hat{x}^{2\alpha}}\right]=\frac{2\pi}{\vartheta_{\alpha}}\delta(k_{2\alpha-1})\delta(k_{2\alpha}) \ .
\end{equation}
Finally, we obtain
\begin{equation}
\label{eq:trace0}
    (2\pi)^{D/2}\mathrm{Pf}(\theta)\mathrm{Tr}[e^{ik_{\mu}\hat{x}^{\mu}}]=(2\pi)^{D}\delta^{(D)}(k) \ ,
\end{equation}
where $\mathrm{Pf}(\theta)=\vartheta_1\vartheta_2\dots \vartheta_{D/2}$.

\section{Quantum field theories on noncommutative spacetime}
\label{app:ncfields}

\subsection{Noncommutative fields}

We choose the space of functions on Euclidean commutative spacetime $\mathbb{R}^D$ such that
\begin{equation}
    \mathrm{sup}_x (1+|x|^2)^{k+n_1+\dots +n_D}|\partial_1^{n_1}\dots \partial_D^{n_D}f(x)|^2< +\infty\ , \qquad \forall k, n_i\in \mathbb{Z}_+ \ .
\end{equation}
For each such $f$, we define the corresponding operator-valued function of the noncommutative coordinates $\hat{x}_{\mu}$
\begin{equation}
\label{eq:what}
    \hat{f}(\hat{x})\equiv\int\frac{d^Dk}{(2\pi)^D}\int d^Dx\ f(x)e^{ik_{\mu}(x^{\mu}-\hat{x}^{\mu})} \ .
\end{equation}
The repeated application of the Baker-Campbell-Hausdorff formula to the exponentials
\begin{equation}
\label{eq:expidentity}
    \prod_{i=1}^n\mathrm{exp}\left(-ik_{i\mu}\hat{x}^{\mu}\right)=\mathrm{exp}\left(-i\sum_{i=1}^nk_{i\mu}\hat{x}^{\mu}\right)\mathrm{exp}\left(-\frac{i}{2}\sum_{i<j}\theta_{\mu\nu}k_i^{\mu}k_j^{\nu}\right) \ ,
\end{equation}
with the aid of the the trace formula in eq. \eqref{eq:trace0}, yields
\begin{equation}
\label{eq:trace}
    (2\pi)^{D/2}\mathrm{Pf}(\theta)\mathrm{Tr}\left[\hat{f}(\hat{x}_1)\ \dots \ \hat{f}(\hat{x}_n)\right]= \int d^Dx\ f_1(x)*\dots * f_n(x) \ , \\ 
\end{equation}
where $*$ denotes the Groenewold-Moyal product
\begin{equation}
    f_1(x)*\dots *f_n(x)=\prod_{j<k}^n\mathrm{exp}\left(\frac{i}{2}\theta^{\mu\nu}\frac{\partial}{\partial x_j^{\mu}}\frac{\partial}{\partial x_k^{\nu}}\right)f_1(x_1)\dots f_n(x_n)\Bigg\rvert_{x_1=\dots =x_n=x} \ ,
\end{equation}
that can be extended at noncoinciding points in an analogous manner. By integrating by parts and using the antisymmetry of the matrix $\theta_{\mu\nu}$, it follows
\begin{align}
	\label{eq:twostarapp}
	\int d^Dx\ f_1(x)*f_2(x)=\int d^Dx\ f_1(x)f_2(x) \ .
\end{align}

\subsection{Noncommutative actions}
\label{app:plan}
\par
We can construct fields on noncommutative spacetime mapping ordinary fields  $\phi$ defined on $\mathbb{R}^D$ into linear operators acting on $\mathcal{H}$, through the map
\begin{equation}
    \phi(x) \longmapsto \hat{\phi}(\hat{x})
\end{equation}
defined in eq. \eqref{eq:what}. The most natural way to construct a scalar action out of these objects is to take the trace of some polynomial $\mathcal{S}$ in $\hat{\phi}$ and its derivatives
\begin{equation}
\label{eq:operator0}
    S=(2\pi)^{D/2}\mathrm{Pf}(\theta)\ \mathrm{Tr}\left[\mathcal{S}\left(\widehat{\partial_{\mu}{\phi}} ,\ \hat{\phi}\right)\right] \ .
\end{equation}
Then, the identity in eq. \eqref{eq:trace} connects the operator to the coordinate representation. For example, the kinetic term of a noncommutative scalar theory reads by eq. \eqref{eq:twostarapp}
\begin{align}
	\label{eq:quadratic}
	S_2=&(2\pi)^{D/2}\mathrm{Pf}(\theta)\ \mathrm{Tr}\left[\frac{1}{2}\hat{\phi}(\hat{x}) \widehat{(-\Box+m^2){\phi}}(\hat{x})\right] \nonumber \\
	=&\frac{1}{2}\int d^Dx\ \phi(x)*\left(-\Box+m^2\right)\phi(x) \nonumber \\
	=&\frac{1}{2}\int d^Dx\ \phi(x)\left(-\Box+m^2\right)\phi(x) \ .
\end{align}
It follows that, in our noncommutative quantum field theory, the quadratic terms in the action and the corresponding propagators are equal to their commutative counterparts
\begin{align}
	\expval{\phi(p) \phi(q)}_0=(2\pi)^D\delta^{(D)}(p+q)\frac{1}{p^2+m^2} \ ,
\end{align}
where $\expval{\dots}_0$ denotes the v.e.v. in the free theory and the Fourier transform $\phi(p)$ of the elementary field $\phi(x)$ is
\begin{align}
	\phi(p)=\int d^Dx\ \phi(x) e^{ip_{\mu}x^{\mu}} \ .
\end{align}
By adding a quartic vertex to eq. \eqref{eq:quadratic}
\begin{equation}
    S_4=(2\pi)^{D/2}\mathrm{Pf}(\theta)\ \mathrm{Tr}\left(\frac{\lambda}{4!} \hat{\phi}^4(\hat{x})\right)=\frac{\lambda}{4!}\int d^Dx\ \phi(x)*\phi(x)*\phi(x)*\phi(x)  \ ,
\end{equation}
the quantization of the corresponding action is performed, as usually, through the Euclidean functional integral. The interaction vertices of the noncommutative fields are easily computed in the momentum representation. In our case
\begin{equation}
\label{eq:ncphi4}
    S_4=\prod_{a=1}^4\int\frac{d^Dk_a}{(2\pi)^D}\phi(k_a)(2\pi)^D\delta^{(D)}\left(\sum_{a=1}^4k_a\right)\left(\prod_{a<b}^4e^{-\frac{i}{2}k_a\wedge k_b}\right) \ ,
\end{equation}
thus implying, contrary to commutative field theories, that the vertices are not symmetric under permutations of external lines, but only under cyclic permutations. 

\subsection{Noncommutativity and planarity}

The aforementioned rigidity of the interaction vertices allows us for a topological classification of Feynman diagrams \emph{à la} 't Hooft, where fields transform under the adjoint representation of the gauge group and interactions are single trace. Perhaps this is not surprising, since the action written in the operator representation in eq. \eqref{eq:ncymop} is manifestly a matrix model, though for practical calculations we employ the coordinate representation in eq. \eqref{eq:ncphi4} to deal with spacetime indices more easily. \par
It follows that in gauge theories where all fields transform under the adjoint representation, the 't Hooft and the noncommutative classification of the topology of Feynman graphs are equivalent and, specifically, the corresponding notion of planarity. \par
\par 
As a consequence \cite{Filk:1996dm}, in a noncommutative field theory a connected planar correlator $\expval{\phi_{i_1}(p_1)\dots \phi_{i_n}(p_n)}^{(\theta)}_{\text{conn, pl}}$ of elementary fields $\phi_i(p)$ in the momentum representation satisfies
\begin{equation}
\label{eq:filk}
    \expval{\phi_{i_1}(p_1)\dots \phi_{i_n}(p_n)}^{(\theta)}_{\text{conn, pl}}=e^{-\frac{i}{2}\sum_{j<j'}p_j\wedge p_{j'}}\expval{\phi_{i_1}(p_1)\dots \phi_{i_n}(p_n)}_{\text{conn, pl}}
\end{equation}
where $\expval{\dots }_{\text{conn, pl}}=\expval{\dots }_{\text{conn, pl}}^{(\theta=0)}$. Thus, for the connected planar graphs in the momentum representation the noncommutativity appears only through an overall $\theta_{\mu\nu}$-dependent phase factor. The corresponding relation in the coordinate representation reads \begin{equation}
\label{eq:filkcoord}
    \expval{\phi_{i_1}(x_1)\dots \phi_{i_n}(x_n)}^{(\theta)}_{\text{conn, pl}}=e^{+\frac{i}{2}\sum_{j<j'}\partial_j\wedge \partial_{j'}}\expval{\phi_{i_1}(x_1)\dots \phi_{i_n}(x_n)}_{\text{conn, pl}} \ .
\end{equation} 
Yet, because nonplanar graphs possess IR divergences that occurr for $\theta_{\mu\nu}\to 0$, in general the commutative limit of noncommutative quantum field theories is not smooth \cite{Minwalla:1999px} beyond the planar limit, even for the correlators of the elementary fields.

\section{Intermediate regularization} \label{C}

We compute the Fourier transform of the regularized propagator in eq. \eqref{eq:regulator} from momentum to coordinate representation in dimension $D$ 
\begin{equation}
    G_D(x;a)=\int\frac{d^Dp}{(2\pi)^D}\frac{1}{p^2}e^{-p^2a^2}e^{ip_{\mu}x^{\mu}}
\end{equation}
by introducing a Schwinger parameter $s$ and performing the Gaussian integral on the momenta
\begin{align}
    G_D(x;a)=&\int\frac{d^Dp}{(2\pi)^D}\int_0^{\infty}dse^{-(s+a^2)p^2+ip_{\mu}x^{\mu}} \nonumber \\
    =&\frac{1}{(4\pi)^{D/2}}\int_0^{\infty}ds\ (s+a^2)^{-D/2}e^{-\frac{x^2}{4(s+a^2)}} \nonumber \\
    =&\frac{1}{(4\pi)^{D/2}}\int_{a^2}^{\infty}ds\ s^{-D/2}e^{-\frac{x^2}{4s}}   \ .
\end{align}
Changing variable to $t=x^2/4s$, we obtain
\begin{align}
    G_D(x;a)=& \frac{1}{4\pi^{D/2}}(x^2)^{1-\frac{D}{2}}\int_{\frac{x^2}{4a^2}}^{\infty}dt\ t^{\frac{D}{2}-2}e^{-t}
\end{align}
that in $D=4$ reduces to
\begin{equation}
    G_{D=4}(x;a)=\frac{1-e^{-\frac{x^2}{4a^2}}}{4\pi^2x^2} \ .
\end{equation}

\bibliographystyle{JHEP}
\bibliography{References}

\end{document}